\documentclass[apjl]{emulateapj}
\usepackage{times}

\shorttitle{The scale dependence of mass assembly in galaxies}
\shortauthors{Mateus et al.}

\newcommand{\msun}{M$_\odot$}

\begin{document}

\title{The scale dependence of mass assembly in galaxies}

\author{Ab\'ilio Mateus\altaffilmark{1,2}, Raul Jimenez\altaffilmark{2,3} and Enrique Gazta\~naga\altaffilmark{2}}

\altaffiltext{1}{Laboratoire d'Astrophysique de Marseille, CNRS UMR6110,  38 rue Fr\'ed\'eric Joliot-Curie, 13388 Marseille, France. abilio.mateus@oamp.fr}
\altaffiltext{2}{Institute of Space Sciences, CSIC-IEEC, Campus UAB, F. de Ci\`encies, Torre C5 par-2, 08193 Barcelona, Spain. gazta@ieec.uab.es}
\altaffiltext{3}{ICREA Professor; also Dept. of Astrophysical Sciences, Princeton University, Princeton, USA. raulj@astro.princeton.edu}

\begin{abstract}
We compare the results of the mark correlation analysis of galaxies in a sample from the Sloan Digital Sky Survey and from two galaxy catalogs obtained by semi-analytical galaxy formation models implemented on the Millennium Simulation. We use the MOPED method to retrieve the star formation history of observed galaxies and use star formation parameters  as weights to the mark correlations. We find an excellent match between models and observations when the mark correlations use stellar mass and luminosity as weights. The most remarkable result is related to the mark correlations associated to the evolution of mass assembly through star formation in galaxies, where we find that semi-analytical models are able to reproduce the main trends seen in the observational data. In addition, we find a good agreement between the redshift evolution of the mean total mass formed by star formation predicted by the models and that measured by MOPED. Our results show that close galaxy pairs today formed more stellar mass $\sim 10$~Gyr ago than the average, while more recently this trend is the opposite, with close pairs showing low levels of star formation activity. We also show a strong correlation in simulations between the shape and time evolution of the star formation marks and the number of major mergers experienced by galaxies, which drive the environmental dependence in galaxy formation by regulating the star formation process.
\end{abstract}


\section{Introduction}

Tools to extract physical information from galaxy spectra have evolved spectacularly in the past five years. It is now possible to extract accurately the past star formation (SF) and metallicity histories of galaxies from their integrated stellar light \citep[e.g.][]{heavens04,cid05,panter07,cid07}. Further, with the advent of mega-spectroscopic surveys, it is now possible to have samples large enough  to study the physical parameters that drive the formation and evolution of galaxies as a function of galaxy properties. One obvious parameter is the galaxy mass, and it seems now firmly established that mass has an influence on the star formation rates (SFRs) of galaxies \citep[e.g.][]{cowie96}. However, and open question is if other parameters besides mass influence the shape and formation of galaxies \citep[e.g.][]{gao05,angulo08}.

In a recent study, \citet{sheth06} used mark correlations of the MOPED SF and metallicities of SDSS galaxies to determine how these quantities were correlated. They found that close pairs ( $ < 1$~Mpc) at $z \sim 0.1$ dominated the SF activity in the past relative to the mean, while today the SF is dominated by pairs that are separated by more than $10$~Mpc. Similar trends were found for stellar metallicity. While this result fits well within the framework of galaxy ``downsizing'' it is unclear how current models of galaxy formation are able, or not, to reproduce it. In fact, a halo occupation distribution (HOD) study shows that the trends found in \citet{sheth06} cannot be reproduced by changing the free parameters in the HOD model, thus pointing toward a need for some environmental dependence, which the current HOD paradigm does not incorporate. Thus, it is useful to look into numerical simulations how environment can influence galaxy evolution.

In this Letter, we explore how semi-analytical models (SAMs) of galaxy formation perform at reproducing the mark correlations computed in \citet{sheth06}. We use the Millennium Simulation and two semi-analytical recipes implemented on it to follow galaxy evolution. We find that models are successful at reproducing the trends observed in the SDSS mark study. The models therefore do well at reproducing the environmental trends of SF suggesting that mass is not the only driver of galaxy formation and evolution.

\section{Star formation history of galaxies}

The observational galaxy sample used in this work was drawn from the 
SDSS Data Release 3 \citep[DR3;][]{abazajian05}.  We built a volume 
limited sample with  $r$-band absolute magnitude $M_r < -20$
and redshift range $0.02 < z < 0.071$, with a conservative limit in the 
apparent magnitude of $m_r \le 17.5$. This sample corresponds to the 
faint catalog used by \cite{sheth06}. We used results obtained by the 
MOPED algorithm, described in detail by \cite{panter07}, to extract star 
 formation histories (SFH) from SDSS galaxy spectra. It uses the stellar 
population models of \cite{BC03} and the \cite{chabrier03}
initial mass function (IMF).

The simulated galaxy catalog was obtained from the Millennium Simulation (MS) 
of dark matter structure growth, which follows the evolution of $N = 2160^3$ particles of mass 
$8.6 \times 10^8 h^{-1}$~\msun, within a comoving box of size $500 h^{-1}$~Mpc on a side, from redshift $z = 127$ to the present \citep{springel05}. The adopted cosmological model is the concordance $\Lambda$CDM model with parameters $\Omega_m = 0.25$, $\Omega_b=0.045$, $h=0.73$, $\Omega_\Lambda=0.75$, $n=1$ and $\sigma_8=0.9$. These parameters are also adopted along this paper. Haloes and subhaloes in 64 output snapshots were identified using the SUBFIND algorithm described in \citet{springel01}, and merger trees were then constructed that describe how haloes grow as the model universe evolves. These merger trees form the basic input needed by the SAMs implemented in the simulation. We used results obtained by the model described by \cite{delucia_blaizot07} (Munich model), which is actually a modified version of models used in \cite{croton06} and \cite{delucia06}, and results from the model described  in detail by \citet{bower06} (Durham model), which is an extension of the GALFORM model implemented by \citet{cole00} and \citet{benson03}. These SAMs have many differences in the overall procedure to form  galaxies from the dark matter halos. Particularly, the build up of merger trees, which form the basis of the SAMs integrated on numerical simulations, has been done in an independent way. Moreover, many aspects of the physics involved in the galaxy formation process are also treated in a different form, mainly those related to the inclusion of AGN feedback to regulate SF in galaxies.

We derived the predicted galaxy mass assembly histories from the SAMs implemented on the MS. In order to compare with our nearby sample of SDSS galaxies, we selected all $z=0$ model galaxies with AB $r$-band absolute magnitudes $M_r < -20$, resulting in two base catalogs containing about $1.15\times10^6$ galaxies from the Durham model and about $1.81\times10^6$ galaxies from the Munich model. For all galaxies in these catalogs, we recovered their complete merger trees by using the post-processing facilities in the MS databases. The mass assembly histories of the model galaxies were then retrieved by summing up the stellar masses of all the progenitors of a galaxy at each time-step of the simulation and the total mass assembled through SF between two time-steps was obtained by taking the difference of the total stellar mass retrieved at each time. Additionally, we have also computed the number of major mergers experienced by a galaxy along its entire life by summing up all merger events with  mass ratio above $1/3$ of the main progenitor stellar mass at each time.

\section{Mark correlations}

\begin{figure}
\plotone{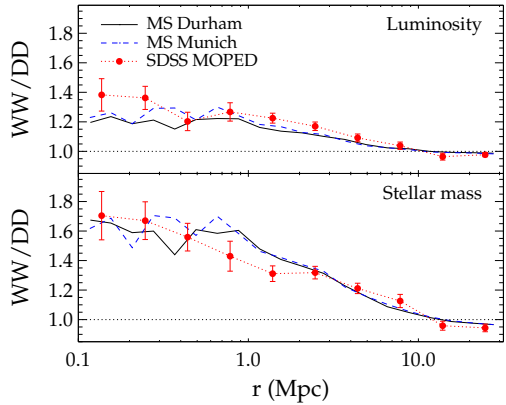}
\caption{Mark correlation analysis for galaxies from the observational SDSS sample (filled circles; dotted lines) and from the Durham model (solid lines) and the Munich model (dashed lines). Error bars for the observational data were obtained with a jackknife procedure. There is an excellent agreement between the observed and the predicted correlations when both $r$-band luminosity (top panel) and stellar mass (bottom panel) are used as the marks.}
\label{fig:1}
\end{figure}

The SFH outputs, as well as other galaxy properties like luminosity and stellar mass, were used to compute the mark correlation functions for both the observational SDSS sample and the MS galaxy catalogs. In our analysis, the mark correlations were computed with the help of the estimator $WW/DD$. Here $DD$ refers to the number of galaxy pairs separated by distances $r \pm dr$ in each bin. While $WW$ is the sum of the products of the galaxy properties weights $W$, also called marks, for  the same pairs as in $DD$. The marks $W$ are normalized so that the mean value (over the galaxies in the sample) is unity. This mark correlation ratio $WW/DD$ is therefore  insensitive to the actual value of the mean clustering (and its evolution).  It is also insensitive to the mean value of the marks or its evolution.  It just indicates if marks are spatially correlated above (or below) the average clustering.

In Fig.~\ref{fig:1} we show a comparison of the mark correlations in redshift-space for the SDSS sample and for the MS galaxy catalogs obtained by the models, where the marks are the $r$-band luminosity and the stellar mass. In the case of the SDSS results, the error bars in this and in subsequent figures were estimated by using a jackknife analysis, following \citet{scranton02}, where we remeasured $WW/DD$ after removing a random region and repeating this procedure 30 times. The comparison between the observed and predicted mark correlations is very good, confirming results obtained by previous studies which show that close pairs tend to be more luminous and massive than the average \cite[see e.g. ][]{skibba06}.
\begin{figure}
\plotone{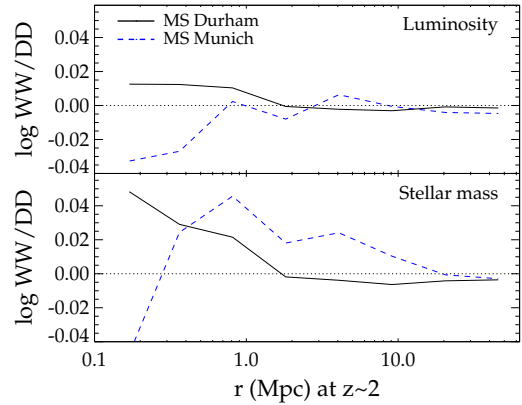}
\caption{Mark correlations for $z=0$ galaxies from the catalogs generated by the SAMs but considering the positions of their main progenitors at $z \sim 2.4$. It is clear from the bottom panel that galaxies that are massive today were more clustered in the past than less massive galaxies.}
\label{fig:2}
\end{figure}

In the MS we can follow the evolution of galaxies by retrieving the information stored in their merger trees. Assuming that the main progenitors represent the present-day galaxies seen at different redshifts, we can investigate in which kind of environment a $z=0$ galaxy inhabited in the past. In Fig.~\ref{fig:2} we do this exercise. We show the redshift-space mark correlations for $z=0$ galaxies in the model catalogs but considering their positions at $z \sim 2.4$. Galaxy pairs containing massive galaxies today were more clustered, on average, at $z \sim 2.4$ than their low-mass counterparts. The same trend is seen for luminosity, but to a less extent, for the Durham model and absent for the Munich model. Thus, in the simulations the relation between stellar mass and environment was already in place more than 10 Gyr ago.

\begin{figure*}
\plotone{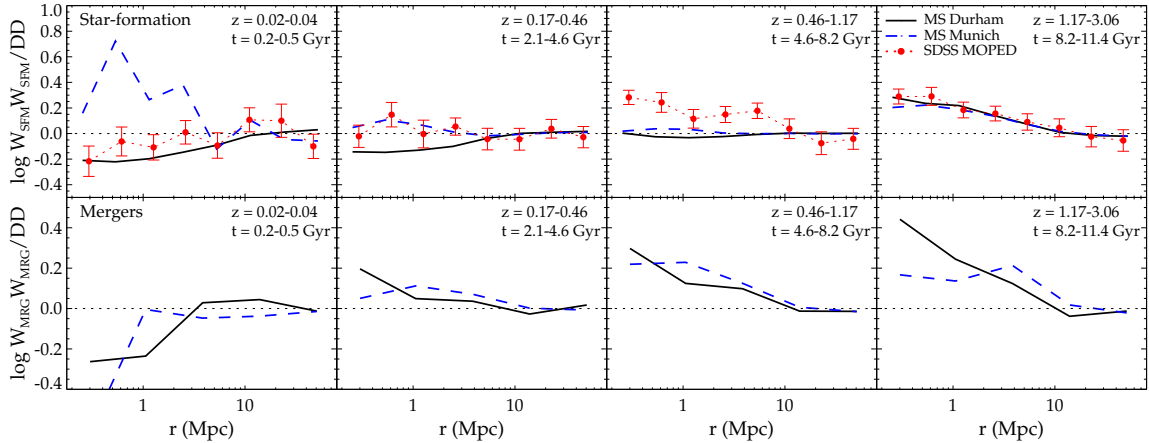}
\caption{Top panels: Evolution of the weighted correlation functions as defined by the total stellar mass formed at different lookback time or redshift bins for the MS using the results from Durham model (solid lines) and Munich model (dashed lines), and for the observational results from the SDSS data using the MOPED (dotted lines; filled circles) approach to retrieve the SFH of galaxies.  Bottom panels: MS merger analysis; the weights are associated to the number of major mergers (with mass ratio larger than $1/3$) that the main progenitor of a $z=0$ galaxy
experienced at each redshift.}
\label{fig:3}
\end{figure*}

On the observational side, the recent advances in the field of spectral synthesis make it possible to retrieve detailed SFH from the spectra of nearby galaxies. Therefore, the evolution of mass assembly in galaxies through SF as predicted by SAMs can be properly compared with results from observational analysis. The mark correlation analysis can be then used to investigate the scale dependence of the SFH of galaxies as predicted by the models and observations. In Fig.~\ref{fig:3}, we show results for both models and the SDSS data. In the top panels of this figure, the mark analysis were done considering the total stellar mass formed (SFM) at a given lookback time bin as the marks. In the case of the SDSS, the trends obtained in this work reproduce the results discussed in \cite{sheth06}. The most interesting result shown in Fig.~\ref{fig:3} is the excellent agreement between the mark correlations obtained by the SAMs,  especially the Durham model, and the observational data. The overall behaviour of the correlations in all lookback time bins shown in this figure indicates that close galaxy pairs as seen at $z=0$ formed more stellar mass at $\sim 10$~Gyr than the average, while more recently this trend is the opposite, with close pairs showing low levels of SF activity in comparison to the average. Note that for the first time bin, the Munich model predicts a positive correlation at smaller scales which is not caused by noise in the data, since the counts are similar to the Durham model. A possible explanation for these trends can be seen in the bottom panels of Fig.~\ref{fig:3}, where we show the mark correlations for the models considering the number of major mergers experienced by a galaxy in a given redshift bin. There is a clear relation between the increment in the SF activity of close  $z=0$ galaxy pairs and the excess of major mergers at $z > 1$, thus reflecting that mergers played a significant role in the build up of galaxies in the high-z Universe.

\section{Conclusions}

In this Letter, we have investigated how the SF in galaxies depends on the scale by comparing the mark correlation analysis of observed and model galaxies. Our results indicate that current semi-analytical galaxy formation models, especially the model by \cite{bower06} (Durham model), implemented on the large Millennium Simulation are able to reproduce the trends seen in the mark correlations obtained for SDSS galaxies, when the marks are derived from the SFH of galaxies (as measured by the MOPED algorithm). Galaxies in close pairs seen nowadays formed stellar mass at a higher rate than the average population at $z\sim2$, and at lower redshifts they show low levels of SF activity compared to the average. Using the predicted mark correlations from the models, we find that for galaxies in $z=0$ close pairs the increase in their SFR at high-redshifts is intimately related to an excess in the number of major merger events they experienced in that epoch. From Fig.~\ref{fig:3} it seems that the measured MOPED SF marks follow closely the marks of the numbers of mergers and seems to suggest that SF follows merger activity. It follows in this scenario that the observed downsizing in SF properties of galaxies was driven by the environment primarily through mergers, which accelerated the evolution of massive galaxies seen today in close pairs (Fig.~\ref{fig:1}). The overall context of these results gives support to a natural path for galaxy evolution which proceeded via a nurture way, with denser environments playing a fundamental role in defining galaxy properties (see detailed discussion in \citealt{mateus07} and \citealt{bundy06}, and references therein).

\begin{figure}
\plotone{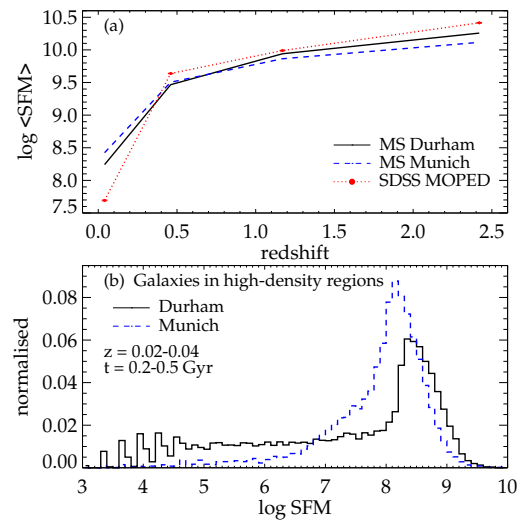}
\caption{(a) Comparison between the logarithm of the mean total stellar mass (SFM; in solar mass) formed in the same redshift bins shown in Fig.~\ref{fig:3} for the models and for the SDSS data. (b) Distribution of the recent SFM for model galaxies located in high-density regions.}
\label{fig:4}
\end{figure}

The correlation properties of SF cannot be reproduced by models that only use the dark halo mass of the galaxy as the only parameter that drives the SF. This result is also related to the evidence of an environmental dependence of halo formation times found by \citet{harker06} and \citet{maulbetsch07}. Actually, current implementations of the HOD paradigm do not include any environmental or spatial correlation. They can therefore not possibly account for the SF effect, which is clearly shown here to be an environmental effect.  It is therefore surprising that current SAMs (especially the Durham one) can reproduce the SF marks so remarkably well, given the complex processes that involve gas-baryon physics. Further, since we had no access to the simulations to fine tune their parameters, the obtained agreement is even more remarkable. The strong correlation between the shape of the SF and the number of mergers mark correlations, which is our main finding,  points toward a possible explanation for regulating the SF activity, which is something that galaxy formation models have inbuilt. However, if this was only the case the Munich models would be similar to the Durham ones. Their main difference resides in the way the merger process is translated to a SF recipe. In order to investigate this issue, in Fig.~\ref{fig:4}a we show the evolution of the mean SFM  in the same redshift bins of Fig.~\ref{fig:3}, where the scale dependence of the fluctuations relative to these mean values was previously shown. Galaxies in the Munich model tend to form more stars at recent times than the Durham model but such difference is not too significative to produce the positive correlation shown in Fig.~\ref{fig:3}. Overall, the model galaxies form less stars at $z > 0.5$ compared to SDSS results and show an excess of SF in the first redshift bin. In Fig.~\ref{fig:4}b we compare the distributions of the recent SFM predicted by the models but now restricting to galaxies located in dense regions. These galaxies have local galaxy density values above the 25th percentile of the distribution of this quantity estimated through a  $k$-nearest neighbor approach (with $k=10$). Galaxies from the Munich model which inhabit dense environments formed 3.3 times more stars in the first time bin shown in Fig.~\ref{fig:3} than their counterparts from the Durham model. This higher recent SF activity predicted by the Munich model for galaxies in high-density regions could be the main factor responsible by the positive correlation at small scales shown in Fig.~\ref{fig:3}. This conclusion is reinforced when we look the SFM distributions for the other time bins, which tend to be similar at high-z, where the mark correlations have almost the same values. Additionally, when we redo the mark analysis for the models considering the same SFM distributions for galaxies in high-density regions, we obtain a lower correlation for the Munich model in the first time bin, confirming that the recent SF activity of galaxies in dense environments predicted by the models is a key factor which defines the differences in their mark correlations. This is also related to the way that models treat energy feedback from stars and AGN and how gas is made available to galaxies to cool and form stars \citep[e.g][]{BDN07,DT08,DB08}. Clearly, without further exploration of parameter space in the SAMs it is difficult to assert which parameter is the culprit of the good fit shown in Fig.~\ref{fig:3}. Current SAMs have no specific environmental dependence beyond a scale of $1$~Mpc besides that of the merger process of dark matter and therefore their success must be linked to the way the merger process is then translated into gas forming stars.

\acknowledgments
We gratefully thank the referee Darren Croton for helping to improve this paper. This work was supported 
by the European Commission's ALFA-II programme through its funding of the Latin-american 
European Network for Astrophysics and Cosmology (LENAC). We warmly thank Andrew Benson 
for enlightening comments on semi-analytical galaxy formation models. 
The present work has been supported by grants from the Spanish Ministerio de 
Educacion y Ciencia, the Spanish CISC and the European Union FP7 program.


\end{document}